\documentclass[a4paper,10pt]{article}
\usepackage[T1]{fontenc}
\usepackage[utf8]{inputenc}
\usepackage[italian,english]{babel}
\usepackage[autostyle,italian=guillemets]{csquotes}
\usepackage{multicol}
\usepackage{microtype}
\usepackage{guit}
\usepackage{booktabs}
\usepackage{graphicx}
\usepackage{mathptmx}
\usepackage{amsfonts}
\usepackage{amssymb}
\usepackage{amsmath}
\usepackage{ifpdf}
\usepackage{dcolumn}
\usepackage{textcomp}
\usepackage{tabularx,array}
\usepackage{ulem}
\usepackage{placeins}
\usepackage{epstopdf}
\usepackage{etex}
\date{}

\title{New CCD Photometric Study of AM Cnc} 
\author{Tasselli, D. \\ TS Corporation Srl - Department of Astronomy and Astrophysics \\ Regione Salamia, 10010 Andrate TO - Italy \\ E-mail:diego.tasselli@tscorporation.org} 

\begin{document}
 
 \maketitle
 \begin{abstract}
 \normalsize We present in this paper the new study of variable star AM Cnc, a short period RRab star, in orther to determine, through the light curve and the physical parameters. The Star were observed for a total of 293 sessions shooting, and exhibits light curve modulation, the so called Blazhko effect with the shortest modulation Period=$0.^{d}559233$ ever observed. We observed this star with the 0,6 mt telescope of the Astronomical Observatory of Andrate (OAA) - To and the result detect small but definite modification in temperature and mean radius of the star itself. All results are compared with previously published literature values and discussed.
 \end{abstract}

{\footnotesize This paper was prepared with the \LaTeX \\}
\begin{multicols}%
{2}

\section{\normalsize Introduction} AM Cnc is a RR Lyr type variable star (RRab type) in the constellation of Cancer, located (R.A. $08^\circ $56' 14.83'', Decl. +$11^\circ $37' 20.4''), and it has an V$_M$=14.529, an orbital period of P= $0.^{d}5580015$, at 257 light years from Earth.  RR Lyrae is a particular type of variables stars with asymmetric light curves (steep ascending branches), periods from 0.3 to 1.2 days, and amplitudes from 0.5 to 2 magnitude in V. The most troublesome is the so called Blazhko effect, the light curve modulation of some fraction of RR Lyrae stars. The Blazhko phenomenon has been known for about a century. Most of the known galactic field RRab stars showing light curve modulation were discovered from photographic observations during the first part of the last century. \\Though the advent of the photoelectric, the CCD era has led to spectacular increase of the photometric precision, there is still a lack of extended high accuracy survey of the modulation properties of the galactic field RR Lyrae stars in the Solar neighborhood. Our objective was to obtain a deep study in the V-band, calibrated to Landolt standards, to update the period of the star.
\section{\normalsize Data} \subsection{\normalsize Observations}
The stars was observed in 2011 March and April (UT) with the Richey-Chretien telescope of the TS Corporation on Andrate (TO) - Italy station, equipped with a CCD camera (FLI EEV2 back illuminated, 2048×2048 pixel mm 0,39 arcsex/pix)
with V filters (Johnson-Kron-Cousins). \\The characteristic of instrument are visible in this paper \cite{Tasselli:2011ug}.\\
Fig. 1 show the identification map for the stars measured and the correspondence between our star numbers and result of analysis data. \\Table 1 show the AM Cnc data by CMC 14 Catalog, Table 2 show the journal book of observations and capture image. We do not indicate the values of the Dark Frame as the CCD is cooled to liquid nitrogen, and therefore there were no shooting Dark.
Preliminary processing of all CCD frames, to apply bias and flat field corrections, was alone with standard routines in the IRAF software package.\\ The magnitudes of stars in the Table 3 and 4 are computed by fitting the position and scale of the PSF to each star image in turn, in order of decreasing brightness. \\The Zero Point of the frame is set during the PSF calculation, thought aperture photometry of the stars used to calculate the PSF.
\subsection{\normalsize Transformations and Reductions}
Instrumental magnitudes for all measured stars were transformed to a standard system using fitting coefficients derived from observations of standard stars whose magnitudes have been well established in earlier studies. Calibration telescope data are visible in this paper \cite{Tasselli:ref1}. \\The linear relationships between the magnitudes r ', J, K of the catalog CMC14 (Sloan DSS filter) and standard sizes V and Rc, are as follows \cite{(Dymock:ref1}: 
\begin{equation}Rc = r' - 0.22 \end{equation} 
\begin{equation}V = 0.6278 ( J - K ) + 0.9947 r' \end{equation} 
\noindent r' are the instrumental magnitudes; J and K are the standard magnitudes for stars. \\ Table 1 show the data for AM Cnc and ref stars from CMD 14 Catalog.
\subsection{\normalsize Photometric Error}
To determine the error in the measurement of values, including the instrumentation and the data catalog, it used the method of the standard deviation. The following table highlights the values obtained in this study: \newline \\
\begin{tabular}{|c|c|}
\hline
\bf Data & \bf Photometric Error \\
\hline
23-mar & 0,21 \\
\hline
24-mar & -0,04 \\
\hline
25-mar & 0,40 \\
\hline
04-apr&  0,59 \\
\hline
\end{tabular} \\ \\
{\textbf{{\scriptsize Photometric error for days}}} \\ \\
\normalsize Figures 3, 4, 5 and 6 show the deviation for each observation session for the V band, highlighting the great alignment's between the instrumental values and the values listed in the catalog for the star.
\subsection{\normalsize Comparison with Previous Studies}
The CCD photometric study of AM Cnc for this work, comparable with other paper study, has been published in Table 5.
\section{\normalsize Data Analysis}
The analysis of the data and their calibration to the international system, are shown by this study in Table 3 and 4. \\Column 1 give the number of observation, Column 2 give the Time of observation in JD, Column 3 give the V magnitudo of Am Cnc measured for the time in column 2, Column 4, 5, 6, 7 and 8 give the V magnitude data photometry for Star Ref, measured of this study.
\subsection{\normalsize The Light Curve diagrams and Period Data}
In Fig. 2 we can show the Light Curve Diagram star for this work.
\includegraphics[width=0.5\textwidth{}]{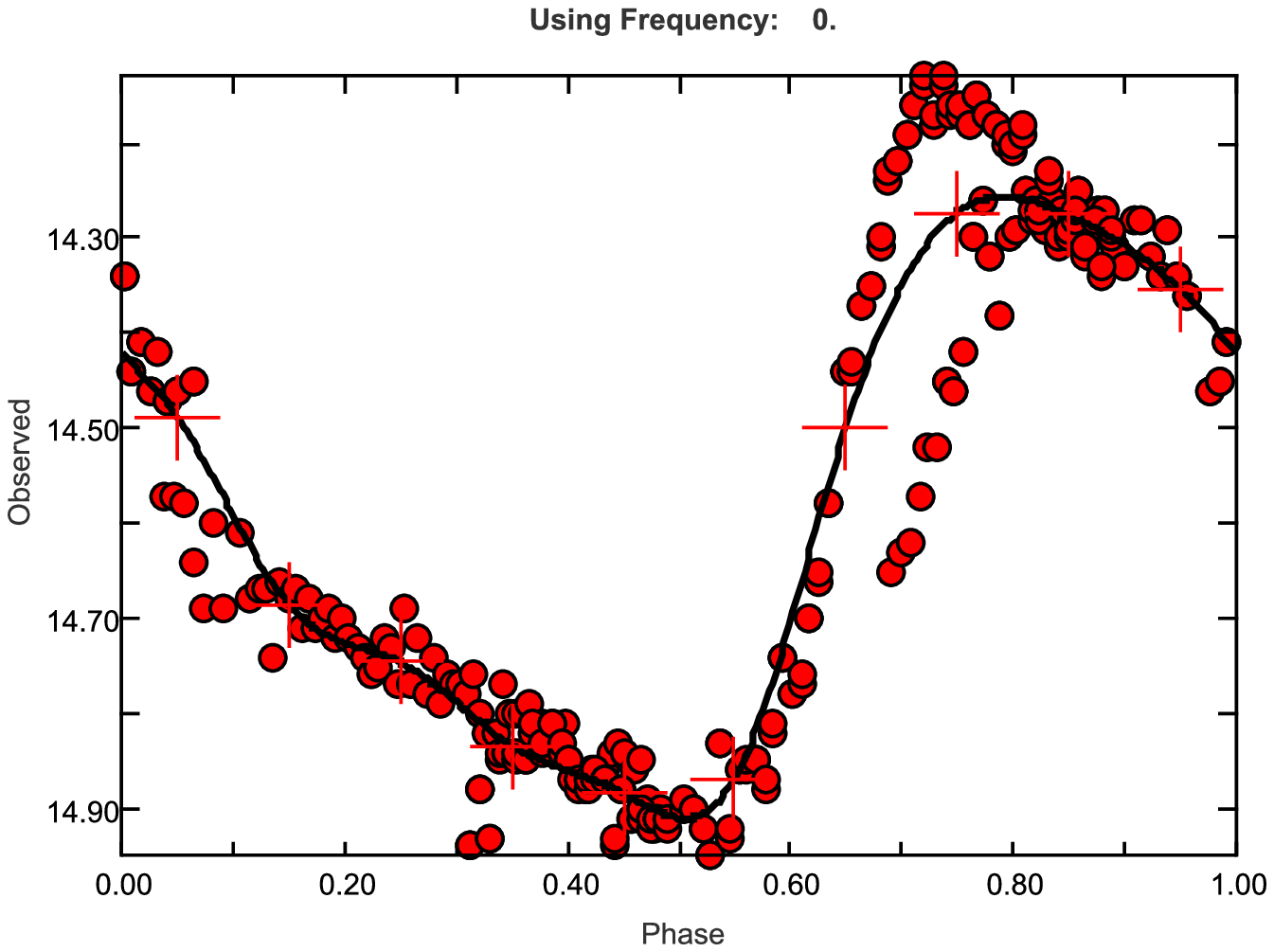} 
\textbf{{\scriptsize Fig 2: V Light Curve Diagram of AM Cnc for this study. The scatter of the light curve is caused by Blazhko modulation. Phase 0.55 is set to the middle of the ascending branch defined as the phase where the V flux is equal to its time averaged value.}} \\ \\ 
Data analysis in this work allowed to obtain the following result for AM Cnc: \\ \\
\begin{tabular}{|c|c|c|}
\hline
{\bf {{\scriptsize Frequence}}} & {\bf {\scriptsize Amplitude}} & {\bf {\scriptsize Phase}} \\
\hline
{\scriptsize \textbf{ 3,40825269e-010} } & {\scriptsize  \textbf{1,847261} }& {\scriptsize \textbf{0,559233} }\\
\hline
\end{tabular} \newline \\ \\
\includegraphics[width=0.5\textwidth{}]{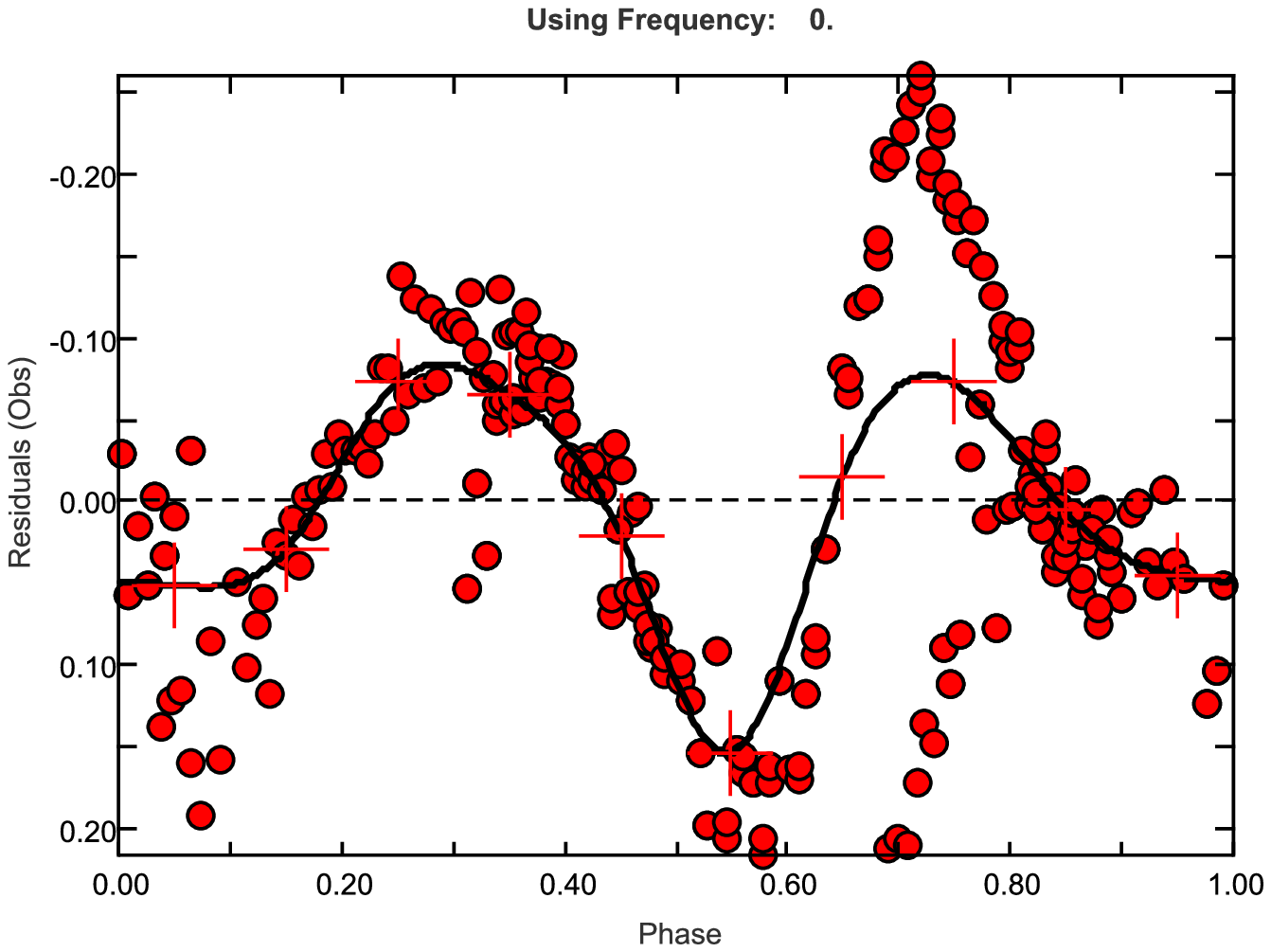} 
\textbf{{\scriptsize Fig 8: Residual Light Curve Diagram for this study}} \\ \\
The residual curve of all the data (Fig. 8) shows that the distortion is symmetrical to phase 0.55 which is set to the phase of the middle
of the rising branch. A similar RR Lyrae, Preston et al. (1965)\cite{Preston:ref2} concluded that a displacement of the shock-forming layer in the atmosphere takes place during the Blazhko cycle. In Preston \& Paczi´nsky (1964)\cite{Preston:ref3} it was also shown that: ``\textit{the phases at which H emission occurs are closely related to phases of photometric parameters}''
Thus the phase on the ascending branch when the visual luminosity of the star equals its time average value (at phase 0.55 in the figures) are closely related, with only some minute differences, to the onset of the H emission. In Figs. 8 the symmetrical modulation is centred exactly on this phase of the pulsation, indicating a connection between the origin of the modulation and that of the H emission.\\
\includegraphics[width=0.5\textwidth{}]{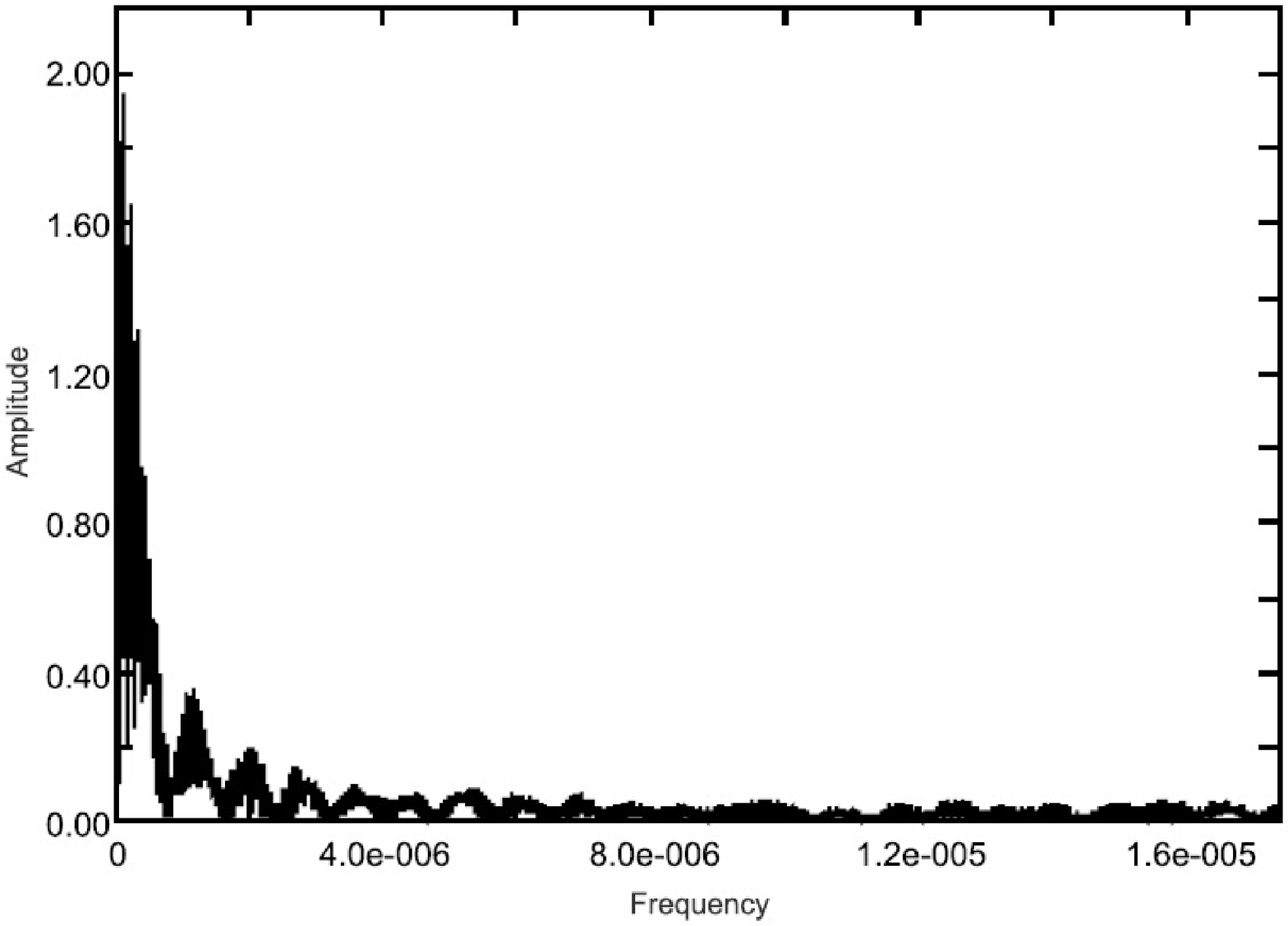}\\
\textbf{{\scriptsize Fig 7: Fourier spectrum of the V light curve of AM Cnc for this study.}}
\subsection{\normalsize Fourier Diagrams}
The Fourier spectra of the V data and the data prewhitened with the pulsation frequency and its harmonics are shown in Fig.7
\section{\normalsize Conclusion}
The study variable AM Cnc, has highlighted the presence Blazhko effect especially highlighted in the diagram for the period of analysis, Fig. 2. The data obtained are also in line with the estimates contained in the data AAVSO.\cite{AAVSO:ref1} \\
The new results concerning the properties of the modulation of AM Cnc are summarized in the next items:
\begin{itemize}
\item The light curve cannot be completely fitted with symmetrically placed Fourier frequency components alone. Residual scatter of the light   curve is still concentrated in the ascending branch, indicating some irregular behaviour of the modulation;
\item The modulation is concentrated in a 0.2 phase interval of the pulsation(Fig.8)
\end{itemize}
Give the importance of this type of stars about Blazhko effect and the still insufficient set of data available, further observations are needed.
\section{\normalsize Acknowledgments}
I would like to thank Dr.ssa Silvia Gargano for supporting me during this study. \\The constructive comments are highly appreciated.
\end{multicols}
\newpage
\begin{figure}
\begin{center}
\includegraphics[width=1\textwidth]{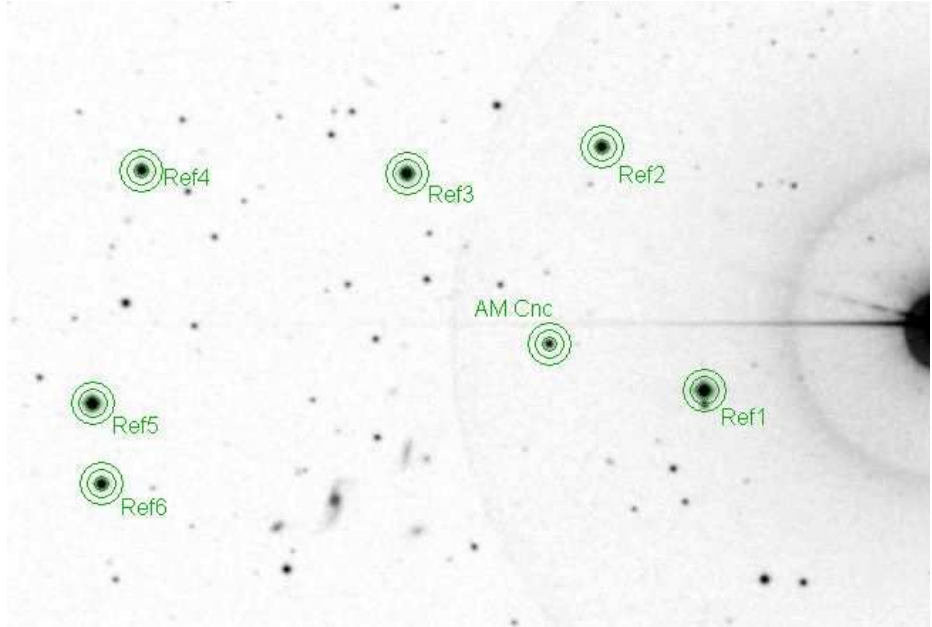}
\caption{Map of the stars AM Cnc. and Ref used in this paper}
\end{center}
\end{figure}
\centering
\newpage
\begin{table}
\caption{The AM CnC data by CMD 14 Catalog} 
\small
\begin{tabular}{|c|c|c|c|c|c|c|c|}
\hline
{\bf Star} &    {\bf r} &    {\bf J} &    {\bf H} &    {\bf K} &    {\bf V} &   {\bf Rc} & {\bf V-Rc} \\
\hline
           &            &            &            &            &            &            &            \\
\hline
    AM Cnc &     14.287 &     13.541 &     13.265 &     13.308 &     14.358 &     14.067 &      0,202 \\
\hline
      Ref1 &     11.496 &     10.528 &     10.215 &     10.140 &     11.679 &     11.276 &      0,280 \\
\hline
      Ref2 &     12.781 &     11.900 &     11.621 &     11.585 &     12.911 &     12.561 &      0,243 \\
\hline
      Ref3 &     12.285 &     11.421 &     11.173 &     11.096 &     12.424 &     12.065 &      0,249 \\
\hline
      Ref4 &     12.982 &     12.143 &     11.916 &     11.908 &     13.061 &     12.762 &      0,208 \\
\hline
      Ref5 &     11.767 &     10.758 &     10.428 &     10.416 &     11.919 &     11.547 &      0,258 \\
\hline
 Ref6/chk1 &     13.006 &     12.064 &     11.822 &     11.807 &     13.098 &     12.786 &      0,217 \\
\hline
\end{tabular}  
\end{table}
\begin{table}
\tiny
\caption{Journal Book of observations and capture image } 
\tiny
\begin{tabular}{|r|r|r|r|r|r|r|r|cccc|}
\hline
\multicolumn{ 1}{|c}{{\bf Data}} & \multicolumn{ 1}{|c}{{\bf Sky}} & \multicolumn{ 1}{|c}{{\bf Seeing}} & \multicolumn{ 1}{|c}{{\bf Moon}} & \multicolumn{ 1}{|c}{{\bf Umidity}} & \multicolumn{ 1}{|c}{{\bf Filter}} & \multicolumn{ 1}{|c}{{\bf Exposition Time}} & \multicolumn{ 1}{|c}{{\bf Open Circles}} &           \multicolumn{ 4}{|c|}{{\bf Calibration}} \\
\multicolumn{ 1}{|c}{{\bf }} & \multicolumn{ 1}{|c}{} & \multicolumn{ 1}{|c}{} & \multicolumn{ 1}{|c}{} & \multicolumn{ 1}{|c}{} & \multicolumn{ 1}{|c}{{\bf }} & \multicolumn{ 1}{|c}{} & \multicolumn{ 1}{|c}{} & \multicolumn{ 2}{|c}{{\bf Dark}} & \multicolumn{ 2}{|c|}{{\bf Flat Field}} \\
\multicolumn{ 1}{|c}{} & \multicolumn{ 1}{|c}{} & \multicolumn{ 1}{|c|}{} & \multicolumn{ 1}{c}{} & \multicolumn{ 1}{|c}{} & \multicolumn{ 1}{|c}{} & \multicolumn{ 1}{|c}{} & \multicolumn{ 1}{|c|}{} & {\bf Number} & {\bf Time of exposition} & {\bf Number} & {\bf Time of exposition} \\
\hline
           &            &            &            &            &            &            &            &            &            &            &            \\
23/03/2011 &      Clear &       9/10 &       84\% &       70\% &          V &    240 sec & 2, 3 and 7 &          - &          - &         30 &          2 \\
24/03/2011 &      Clear &       7/10 &       76\% &       50\% &          V &    240 sec & 2, 3 and 7 &          - &          - &         30 &          2 \\
25/03/2011 &      Clear &       7/10 &       66\% &       90\% &          V &    240 sec & 2, 3 and 7 &          - &          - &         30 &          2 \\
04/04/2011 &      Clear &       9/10 &        1\% &       90\% &          V &    240 sec & 2, 3 and 7 &          - &          - &         30 &          2 \\
\hline
\end{tabular}  
\end{table}
\begin{table}
\tiny
\caption{V Magnitudo Data for Am Cnc and Ref stars} 
\tiny
\begin{tabular}{|c|c|c|c|c|c|c|c||c|c|c|c|c|c|c|c|}
\hline
  {\bf ID} & {\bf T (JD)} & {\bf Obj1} & {\bf Ref1} & {\bf Ref2} & {\bf Ref3} & {\bf Ref4} & {\bf Ref5} &   {\bf ID} & {\bf T (JD)} & {\bf Obj1} & {\bf Ref1} & {\bf Ref2} & {\bf Ref3} & {\bf Ref4} & {\bf Ref5} \\
\hline
         1 & 24556442887499900 &      14,57 &      11,93 &      12,47 &      12,98 &      11,69 &      12,32 &         86 & 24556463748198400 &      14,26 &      11,65 &      12,94 &      12,44 &      13,17 &      11,90 \\
\hline
         2 & 24556442918981400 &      14,57 &      11,93 &      12,47 &      12,98 &      11,69 &      12,32 &         87 & 24556463776962100 &      14,27 &      11,65 &      12,94 &      12,45 &      13,16 &      11,90 \\
\hline
         3 & 24556442950231400 &      14,58 &      11,94 &      12,47 &      12,98 &      11,68 &      12,32 &         88 & 24556463805736300 &      14,28 &      11,65 &      12,94 &      12,44 &      13,14 &      11,89 \\
\hline
         4 & 24556442981597200 &      14,64 &      11,93 &      12,47 &      12,99 &      11,68 &      12,32 &         89 & 24556463834475500 &      14,25 &      11,66 &      12,91 &      12,45 &      13,15 &      11,90 \\
\hline
         5 & 24556443012847200 &      14,69 &      11,93 &      12,48 &      12,98 &      11,68 &      12,32 &         90 & 24556463863246000 &      14,29 &      11,65 &      12,95 &      12,45 &      13,15 &      11,90 \\
\hline
         6 & 24556443044212900 &      14,60 &      11,94 &      12,47 &      12,97 &      11,69 &      12,32 &         91 & 24556463892091600 &      14,27 &      11,66 &      12,95 &      12,45 &      13,16 &      11,90 \\
\hline
         7 & 24556443075462900 &      14,69 &      11,94 &      12,47 &      12,98 &      11,69 &      12,31 &         92 & 24556463920827400 &      14,27 &      11,64 &      12,93 &      12,46 &      13,17 &      11,89 \\
\hline
         8 & 24556443126504600 &      14,61 &      11,94 &      12,46 &      12,99 &      11,69 &      12,32 &         93 & 24556463949646600 &      14,31 &      11,66 &      12,94 &      12,44 &      13,14 &      11,90 \\
\hline
         9 & 24556443157870300 &      14,68 &      11,93 &      12,48 &      12,99 &      11,68 &      12,31 &         94 & 24556463978431000 &      14,33 &      11,66 &      12,95 &      12,46 &      13,16 &      11,89 \\
\hline
        10 & 24556443189120300 &      14,67 &      11,93 &      12,47 &      12,99 &      11,68 &      12,31 &         95 & 24556464007197800 &      14,28 &      11,66 &      12,93 &      12,44 &      13,17 &      11,91 \\
\hline
        11 & 24556443216319400 &      14,67 &      11,94 &      12,48 &      12,98 &      11,68 &      12,32 &         96 & 24556464036027400 &      14,28 &      11,65 &      12,93 &      12,47 &      13,13 &      11,90 \\
\hline
        12 & 24556443238657400 &      14,74 &      11,94 &      12,47 &      12,99 &      11,68 &      12,31 &         97 & 24556464064780500 &      14,32 &      11,65 &      12,95 &      12,47 &      13,13 &      11,91 \\
\hline
        13 & 24556443260879600 &      14,66 &      11,94 &      12,47 &      12,99 &      11,68 &      12,31 &         98 & 24556464093585800 &      14,34 &      11,67 &      12,94 &      12,44 &      13,13 &      11,89 \\
\hline
        14 & 24556443283217500 &      14,68 &      11,94 &      12,48 &      12,98 &      11,68 &      12,30 &         99 & 24556464122369000 &      14,29 &      11,66 &      12,96 &      12,45 &      13,14 &      11,90 \\
\hline
        15 & 24556443305324000 &      14,67 &      11,94 &      12,47 &      12,98 &      11,68 &      12,32 &        100 & 24556464151158100 &      14,34 &      11,64 &      12,96 &      12,43 &      13,13 &      11,90 \\
\hline
        16 & 24556443327546200 &      14,71 &      11,94 &      12,48 &      12,98 &      11,69 &      12,31 &        101 & 24556464179974900 &      14,36 &      11,65 &      12,95 &      12,45 &      13,14 &      11,90 \\
\hline
        17 & 24556443349768500 &      14,68 &      11,94 &      12,47 &      12,98 &      11,68 &      12,32 &        102 & 24556464255725400 &      14,46 &      11,64 &      12,97 &      12,43 &      13,18 &      11,91 \\
\hline
        18 & 24556443372222200 &      14,71 &      11,95 &      12,48 &      12,98 &      11,68 &      12,31 &        103 & 24556464284522600 &      14,45 &      11,65 &      12,94 &      12,44 &      13,19 &      11,91 \\
\hline
        19 & 24556443394560100 &      14,70 &      11,93 &      12,47 &      12,99 &      11,69 &      12,31 &        104 & 24556464313343400 &      14,41 &      11,66 &      12,94 &      12,44 &      13,11 &      11,91 \\
\hline
        20 & 24556443416782400 &      14,69 &      11,94 &      12,47 &      13,00 &      11,68 &      12,31 &        105 & 24556464342094200 &      14,34 &      11,65 &      12,94 &      12,44 &      13,16 &      11,91 \\
\hline
        21 & 24556443439351800 &      14,72 &      11,94 &      12,47 &      12,98 &      11,68 &      12,32 &        106 & 24556464370820700 &      14,44 &      11,67 &      12,94 &      12,42 &      13,17 &      11,90 \\
\hline
        22 & 24556443461458300 &      14,70 &      11,94 &      12,48 &      12,97 &      11,68 &      12,31 &        107 & 24556464399573800 &      14,41 &      11,65 &      12,94 &      12,44 &      13,15 &      11,90 \\
\hline
        23 & 24556443483912000 &      14,72 &      11,94 &      12,47 &      12,98 &      11,68 &      12,31 &        108 & 24556464428343500 &      14,46 &      11,66 &      12,94 &      12,43 &      13,15 &      11,90 \\
\hline
        24 & 24556443506481400 &      14,73 &      11,94 &      12,47 &      12,99 &      11,68 &      12,32 &        109 & 24556464457294900 &      14,42 &      11,65 &      12,94 &      12,43 &      13,13 &      11,90 \\
\hline
        25 & 24556443528703700 &      14,74 &      11,94 &      12,47 &      12,98 &      11,69 &      12,31 &        110 & 24556464486107100 &      14,47 &      11,65 &      12,94 &      12,46 &      13,14 &      11,90 \\
\hline
        26 & 24556443551157400 &      14,76 &      11,94 &      12,47 &      12,99 &      11,67 &      12,31 &        111 & 24556464514906700 &      14,46 &      11,66 &      12,95 &      12,45 &      13,13 &      11,89 \\
\hline
        27 & 24556443573726800 &      14,75 &      11,94 &      12,47 &      12,98 &      11,68 &      12,32 &        112 & 24556464572396800 &      14,45 &      11,64 &      12,93 &      12,43 &      13,16 &      11,91 \\
\hline
        28 & 24556443596064800 &      14,72 &      11,94 &      12,47 &      12,97 &      11,69 &      12,32 &        113 & 24556562624475600 &      14,94 &      11,64 &      12,93 &      12,46 &      13,17 &      11,90 \\
\hline
        29 & 24556443618402700 &      14,73 &      11,94 &      12,48 &      12,98 &      11,68 &      12,31 &        114 & 24556562653296700 &      14,88 &      11,63 &      12,94 &      12,45 &       3,16 &      11,91 \\
\hline
        30 & 24556443640509200 &      14,77 &      11,94 &      12,48 &      12,99 &      11,68 &      12,31 &        115 & 24556562682046400 &      14,93 &      11,64 &      12,93 &      12,46 &      13,15 &      11,91 \\
\hline
        31 & 24556443662847200 &      14,69 &      11,94 &      12,47 &      12,98 &      11,68 &      12,31 &        116 & 24556562710837800 &      14,85 &      11,64 &      12,94 &      12,46 &      13,17 &      11,91 \\
\hline
        32 & 24556443685185100 &      14,77 &      11,94 &      12,47 &      12,98 &      11,68 &      12,32 &        117 & 24556562739659100 &      14,84 &      11,64 &      12,95 &      12,46 &      13,16 &      11,90 \\
\hline
        33 & 24556443707407400 &      14,72 &      11,93 &      12,47 &      12,98 &      11,69 &      12,32 &        118 & 24556562768391400 &      14,85 &      11,64 &      12,94 &      12,46 &      13,16 &      11,90 \\
\hline
        34 & 24556443729745300 &      14,78 &      11,94 &      12,48 &      12,97 &      11,68 &      12,31 &        119 & 24556562797152600 &      14,85 &      11,64 &      12,95 &      12,46 &      13,17 &      11,90 \\
\hline
        35 & 24556443752314800 &      14,74 &      11,94 &      12,47 &      12,99 &      11,68 &      12,32 &        120 & 24556562825894200 &      14,82 &      11,64 &      12,94 &      12,45 &      13,16 &      11,91 \\
\hline
        36 & 24556443774652700 &      14,79 &      11,93 &      12,48 &      12,98 &      11,69 &      12,31 &        121 & 24556562854765000 &      14,84 &      11,64 &      12,95 &      12,46 &      13,16 &      11,90 \\
\hline
        37 & 24556443796874900 &      14,76 &      11,94 &      12,47 &      12,99 &      11,68 &      12,31 &        122 & 24556562883505500 &      14,81 &      11,64 &      12,94 &      12,45 &      13,14 &      11,91 \\
\hline
        38 & 24556443819097200 &      14,77 &      11,94 &      12,47 &      12,98 &      11,69 &      12,31 &        123 & 24556562912302600 &      14,84 &      11,64 &      12,93 &      12,45 &      13,16 &      11,91 \\
\hline
        39 & 24556443841319400 &      14,77 &      11,94 &      12,47 &      12,97 &      11,68 &      12,32 &        124 & 24556562941062800 &      14,85 &      11,64 &      12,95 &      12,45 &      13,14 &      11,91 \\
\hline
        40 & 24556443863541600 &      14,78 &      11,93 &      12,47 &      12,98 &      11,69 &      12,31 &        125 & 24556562969867800 &      14,88 &      11,64 &      12,94 &      12,46 &      13,17 &      11,90 \\
\hline
        41 & 24556443885648100 &      14,76 &      11,94 &      12,47 &      12,98 &      11,68 &      12,32 &        126 & 24556562998640600 &      14,88 &      11,64 &      12,97 &      12,46 &      13,12 &      11,91 \\
\hline
        42 & 24556443907870300 &      14,80 &      11,94 &      12,47 &      12,98 &      11,68 &      12,32 &        127 & 24556563027421600 &      14,87 &      11,64 &      12,94 &      12,46 &      13,15 &      11,90 \\
\hline
        43 & 24556443930092500 &      14,82 &      11,94 &      12,47 &      12,99 &      11,68 &      12,31 &        128 & 24556563056262800 &      14,87 &      11,64 &      12,97 &      12,46 &      13,09 &      11,92 \\
\hline
        44 & 24556443952430500 &      14,82 &      11,94 &      12,47 &      12,98 &      11,68 &      12,32 &        129 & 24556563085015900 &      14,94 &      11,64 &      12,96 &      12,45 &      13,12 &      11,91 \\
\hline
        45 & 24556443974652700 &      14,77 &      11,94 &      12,47 &      12,97 &      11,68 &      12,32 &        130 & 24556563113772500 &      14,88 &      11,64 &      12,95 &      12,47 &      13,11 &      11,91 \\
\hline
        46 & 24556443997106400 &      14,80 &      11,93 &      12,47 &      12,99 &      11,68 &      12,31 &        131 & 24556563142537200 &      14,91 &      11,64 &      12,95 &      12,47 &      13,10 &      11,91 \\
\hline
        47 & 24556444019328700 &      14,80 &      11,94 &      12,47 &      12,98 &      11,69 &      12,32 &        132 & 24556563171347400 &      14,91 &      11,64 &      12,95 &      12,47 &      13,13 &      11,90 \\
\hline
        48 & 24556444041666600 &      14,80 &      11,94 &      12,47 &      12,98 &      11,68 &      12,32 &        133 & 24556563200144600 &      14,92 &      11,63 &      12,97 &      12,47 &      13,12 &      11,91 \\
\hline
        49 & 24556444063888800 &      14,79 &      11,94 &      12,48 &      12,97 &      11,69 &      12,32 &        134 & 24556563228940700 &      14,91 &      11,64 &      12,96 &      12,47 &      13,11 &      11,91 \\
\hline
        50 & 24556444086111100 &      14,83 &      11,94 &      12,47 &      12,98 &      11,68 &      12,32 &        135 & 24556563257759900 &      14,92 &      11,64 &      12,95 &      12,47 &      13,12 &      11,91 \\
\hline
        51 & 24556444108333300 &      14,81 &      11,94 &      12,47 &      12,98 &      11,68 &      12,32 &        136 & 24556563313971700 &      14,90 &      11,64 &      12,94 &      12,45 &      13,18 &      11,90 \\
\hline
        52 & 24556444130787000 &      14,83 &      11,94 &      12,48 &      12,98 &      11,68 &      12,31 &        137 & 24556563342824500 &      14,90 &      11,64 &      12,94 &      12,45 &      13,16 &      11,91 \\
\hline
        53 & 24556444153009200 &      14,83 &      11,94 &      12,47 &      12,98 &      11,68 &      12,32 &        138 & 24556563371648700 &      14,92 &      11,64 &      12,94 &      12,45 &      13,18 &      11,90 \\
\hline
        54 & 24556444175347200 &      14,81 &      11,94 &      12,47 &      12,98 &      11,68 &      12,32 &        139 & 24556563400458600 &      14,95 &      11,64 &      12,95 &      12,44 &      13,18 &      11,90 \\
\hline
        55 & 24556444197685100 &      14,87 &      11,94 &      12,48 &      12,97 &      11,68 &      12,32 &        140 & 24556563429281300 &      14,83 &      11,64 &      12,94 &      12,46 &      13,17 &      11,90 \\
\hline
        56 & 24556444242245300 &      14,88 &      11,94 &      12,48 &      12,97 &      11,69 &      12,32 &        141 & 24556563458080700 &      14,93 &      11,64 &      12,94 &      12,45 &      13,18 &      11,90 \\
\hline
        57 & 24556444264467500 &      14,86 &      11,94 &      12,48 &      12,98 &      11,68 &      12,32 &        142 & 24556563487014600 &      14,86 &      11,64 &      12,95 &      12,44 &      13,18 &      11,91 \\
\hline
        58 & 24556444286805500 &      14,86 &      11,94 &      12,47 &      12,98 &      11,68 &      12,32 &        143 & 24556563515848900 &      14,86 &      11,65 &      12,94 &      12,45 &      13,17 &      11,90 \\
\hline
        59 & 24556444309143500 &      14,86 &      11,94 &      12,48 &      12,98 &      11,68 &      12,31 &        144 & 24556563544656500 &      14,85 &      11,65 &      12,95 &      12,45 &      13,16 &      11,89 \\
\hline
        60 & 24556444331712900 &      14,84 &      11,93 &      12,48 &      12,97 &      11,69 &      12,32 &        145 & 24556563573489600 &      14,88 &      11,64 &      12,95 &      12,45 &      13,18 &      11,90 \\
\hline
        61 & 24556444353935100 &      14,83 &      11,93 &      12,48 &      12,98 &      11,68 &      12,32 &        146 & 24556563602315200 &      14,82 &      11,64 &      12,94 &      12,45 &      13,16 &      11,90 \\
\hline
        62 & 24556444376504600 &      14,84 &      11,94 &      12,47 &      12,98 &      11,68 &      12,32 &        147 & 24556563631170300 &      14,74 &      11,64 &      12,96 &      12,45 &      13,17 &      11,90 \\
\hline
        63 & 24556444398958300 &      14,86 &      11,93 &      12,48 &      12,98 &      11,68 &      12,32 &        148 & 24556563660010500 &      14,78 &      11,63 &      12,94 &      12,45 &      13,19 &      11,91 \\
\hline
        64 & 24556444421180500 &      14,85 &      11,94 &      12,47 &      12,99 &      11,67 &      12,32 &        149 & 24556563689034300 &      14,77 &      11,63 &      12,95 &      12,46 &      13,17 &      11,90 \\
\hline
        65 & 24556444443518500 &      14,89 &      11,94 &      12,48 &      12,99 &      11,68 &      12,31 &        150 & 24556563717838500 &      14,70 &      11,64 &      12,97 &      12,46 &      13,18 &      11,89 \\
\hline
        66 & 24556444465740700 &      14,92 &      11,93 &      12,48 &      12,98 &      11,68 &      12,32 &        151 & 24556563746656500 &      14,66 &      11,63 &      12,97 &      12,46 &      13,19 &      11,89 \\
\hline
        67 & 24556444488078700 &      14,90 &      11,93 &      12,47 &      12,99 &      11,68 &      12,31 &        152 & 24556563775566200 &      14,58 &      11,64 &      12,95 &      12,45 &      13,21 &      11,89 \\
\hline
        68 & 24556463230125500 &      14,65 &      11,64 &      12,95 &      12,44 &      13,16 &      11,91 &        153 & 24556563828103000 &      14,44 &      11,64 &      12,94 &      12,46 &      13,16 &      11,90 \\
\hline
        69 & 24556463258864800 &      14,63 &      11,64 &      12,94 &      12,44 &      13,16 &      11,90 &        154 & 24556563856946600 &      14,44 &      11,64 &      12,93 &      12,46 &      13,17 &      11,90 \\
\hline
        70 & 24556463287678900 &      14,62 &      11,65 &      12,97 &      12,43 &      13,16 &      11,90 &        155 & 24556563885721700 &      14,37 &      11,64 &      12,97 &      12,45 &      13,17 &      11,89 \\
\hline
        71 & 24556463316466800 &      14,57 &      11,65 &      12,95 &      12,45 &      13,15 &      11,91 &        156 & 24556563914558400 &      14,35 &      11,64 &      12,96 &      12,47 &      13,16 &      11,89 \\
\hline
        72 & 24556463345250200 &      14,52 &      11,64 &      12,93 &      12,43 &      13,15 &      11,91 &        157 & 24556563943384600 &      14,31 &      11,64 &      12,96 &      12,47 &      13,17 &      11,89 \\
\hline
        73 & 24556463374090500 &      14,52 &      11,65 &      12,92 &      12,45 &      13,16 &      11,91 &        158 & 24556563972191000 &      14,24 &      11,64 &      12,96 &      12,45 &      13,16 &      11,89 \\
\hline
        74 & 24556463402833200 &      14,45 &      11,65 &      12,92 &      12,43 &      13,15 &      11,91 &        159 & 24556564001077500 &      14,22 &      11,64 &      12,94 &      12,48 &      13,16 &      11,90 \\
\hline
        75 & 24556463431592200 &      14,46 &      11,64 &      12,93 &      12,43 &      13,15 &      11,90 &        160 & 24556564029828300 &      14,19 &      11,63 &      12,95 &      12,48 &      13,17 &      11,89 \\
\hline
        76 & 24556463460320900 &      14,42 &      11,65 &      12,94 &      12,44 &      13,16 &      11,90 &        161 & 24556564058639400 &      14,16 &      11,64 &      12,95 &      12,46 &      13,17 &      11,90 \\
\hline
        77 & 24556463489104600 &      14,30 &      11,67 &      12,92 &      12,42 &      13,14 &      11,91 &        162 & 24556564087514200 &      14,14 &      11,63 &      12,93 &      12,47 &      13,18 &      11,90 \\
\hline
        78 & 24556463517924900 &      14,26 &      11,65 &      12,90 &      12,47 &      13,18 &      11,92 &        163 & 24556564116356700 &      14,18 &      11,64 &      12,95 &      12,46 &      13,17 &      11,89 \\
\hline
        79 & 24556463546732600 &      14,32 &      11,65 &      12,94 &      12,43 &      13,18 &      11,91 &        164 & 24556564145191000 &      14,14 &      11,64 &      12,95 &      12,46 &      13,19 &      11,88 \\
\hline
        80 & 24556463575529000 &      14,38 &      11,65 &      12,93 &      12,44 &      13,17 &      11,90 &        165 & 24556564174032200 &      14,17 &      11,65 &      12,92 &      12,47 &      13,17 &      11,89 \\
\hline
        81 & 24556463604287900 &      14,30 &      11,66 &      12,94 &      12,43 &      13,14 &      11,92 &        166 & 24556564202928000 &      14,17 &      11,64 &      12,94 &      12,46 &      13,17 &      11,90 \\
\hline
        82 & 24556463633050400 &      14,29 &      11,64 &      12,94 &      12,45 &      13,17 &      11,92 &        167 & 24556564231788100 &      14,18 &      11,65 &      12,95 &      12,46 &      13,18 &      11,88 \\
\hline
        83 & 24556463661844000 &      14,25 &      11,65 &      12,94 &      12,44 &      13,17 &      11,90 &        168 & 24556564260599300 &      14,15 &      11,65 &      12,95 &      12,46 &      13,17 &      11,88 \\
\hline
        84 & 24556463690645800 &      14,26 &      11,66 &      12,95 &      12,45 &      13,17 &      11,90 &        169 & 24556564289420800 &      14,17 &      11,65 &      12,94 &      12,46 &      13,19 &      11,88 \\
\hline
        85 & 24556463719417500 &      14,29 &      11,66 &      12,95 &      12,44 &      13,16 &      11,91 &        170 & 24556564318225100 &      14,18 &      11,64 &      12,95 &      12,47 &      13,16 &      11,89 \\
\hline
\end{tabular}  
\end{table}
\begin{table}
\tiny
\caption{V Magnitudo Data for Am Cnc and Ref stars} 
\tiny
\begin{tabular}{|c|c|c|c|c|c|c|c||c|c|c|c|c|c|c|c|}
\hline
  {\bf ID} & {\bf T (JD)} & {\bf Obj1} & {\bf Ref1} & {\bf Ref2} & {\bf Ref3} & {\bf Ref4} & {\bf Ref5} &   {\bf ID} & {\bf T (JD)} & {\bf Obj1} & {\bf Ref1} & {\bf Ref2} & {\bf Ref3} & {\bf Ref4} & {\bf Ref5} \\
\hline
171 & 24556564347026900 &      14,20 &      11,65 &      12,94 &      12,47 &      13,18 &      11,87 &        233 & 24556562912302600 &       14,8 &       11,6 &       12,9 &       12,4 &       13,2 &       11,9 \\
\hline
       172 & 24556564375926100 &      14,21 &      11,65 &      12,97 &      12,47 &      13,18 &      11,87 &        234 & 24556562941062800 &       14,9 &       11,6 &         13 &       12,5 &       13,1 &       11,9 \\
\hline
       173 & 24556564404780000 &      14,19 &      11,64 &      12,95 &      12,46 &      13,18 &      11,89 &        235 & 24556562969867800 &       14,9 &       11,6 &       12,9 &       12,5 &       13,2 &       11,9 \\
\hline
       174 & 24556564433609700 &      14,28 &      11,64 &      12,96 &      12,47 &      13,18 &      11,87 &        236 & 24556562998640600 &       14,9 &       11,6 &         13 &       12,5 &       13,1 &       11,9 \\
\hline
       175 & 24556564462423100 &      14,28 &      11,66 &      12,94 &      12,46 &      13,19 &      11,87 &        237 & 24556563027421600 &       14,9 &       11,6 &       12,9 &       12,5 &       13,2 &       11,9 \\
\hline
       176 & 24556564491219200 &      14,24 &      11,65 &      12,94 &      12,47 &      13,16 &      11,88 &        238 & 24556563056262800 &       14,9 &       11,6 &         13 &       12,5 &       13,1 &       11,9 \\
\hline
       177 & 24556564520023300 &      14,31 &      11,64 &      12,99 &      12,47 &      13,16 &      11,87 &        239 & 24556563085015900 &       14,9 &       11,6 &         13 &       12,5 &       13,1 &       11,9 \\
\hline
       178 & 24556564548884200 &      14,30 &      11,65 &      12,95 &      12,46 &      13,18 &      11,88 &        240 & 24556563113772500 &       14,9 &       11,6 &         13 &       12,5 &       13,1 &       11,9 \\
\hline
       179 & 24556564577710500 &      14,28 &      11,65 &      12,95 &      12,46 &      13,20 &      11,88 &        241 & 24556563142537200 &       14,9 &       11,6 &         13 &       12,5 &       13,1 &       11,9 \\
\hline
       180 & 24556564606509900 &      14,32 &      11,64 &      12,97 &      12,48 &      13,20 &      11,87 &        242 & 24556563171347400 &       14,9 &       11,6 &         13 &       12,5 &       13,1 &       11,9 \\
\hline
       181 & 24556564635374700 &      14,28 &      11,65 &      12,94 &      12,46 &      13,20 &      11,88 &        243 & 24556563200144600 &       14,9 &       11,6 &         13 &       12,5 &       13,1 &       11,9 \\
\hline
       182 & 24556564664202000 &      14,34 &      11,65 &      12,95 &      12,48 &      13,18 &      11,87 &        244 & 24556563228940700 &       14,9 &       11,6 &         13 &       12,5 &       13,1 &       11,9 \\
\hline
       183 & 24556564693538100 &      14,30 &      11,64 &      12,95 &      12,48 &      13,19 &      11,88 &        245 & 24556563257759900 &       14,9 &       11,6 &       12,9 &       12,5 &       13,1 &       11,9 \\
\hline
       184 & 24556463402833200 &      14,45 &      11,65 &      12,92 &      12,43 &      13,15 &      11,91 &        246 & 24556563313971700 &       14,9 &       11,6 &       12,9 &       12,5 &       13,2 &       11,9 \\
\hline
       185 & 24556463431592200 &      14,46 &      11,64 &      12,93 &      12,43 &      13,15 &      11,90 &        247 & 24556563342824500 &       14,9 &       11,6 &       12,9 &       12,5 &       13,2 &       11,9 \\
\hline
       186 & 24556463460320900 &      14,42 &      11,65 &      12,94 &      12,44 &      13,16 &      11,90 &        248 & 24556563371648700 &       14,9 &       11,6 &       12,9 &       12,5 &       13,2 &       11,9 \\
\hline
       187 & 24556463489104600 &      14,30 &      11,67 &      12,92 &      12,42 &      13,14 &      11,91 &        249 & 24556563400458600 &         15 &       11,6 &         13 &       12,4 &       13,2 &       11,9 \\
\hline
       188 & 24556463517924900 &      14,26 &      11,65 &      12,90 &      12,47 &      13,18 &      11,92 &        250 & 24556563429281300 &       14,8 &       11,6 &       12,9 &       12,5 &       13,2 &       11,9 \\
\hline
       189 & 24556463546732600 &      14,32 &      11,65 &      12,94 &      12,43 &      13,18 &      11,91 &        251 & 24556563458080700 &       14,9 &       11,6 &       12,9 &       12,4 &       13,2 &       11,9 \\
\hline
       190 & 24556463575529000 &      14,38 &      11,65 &      12,93 &      12,44 &      13,17 &      11,90 &        252 & 24556563487014600 &       14,9 &       11,6 &       12,9 &       12,4 &       13,2 &       11,9 \\
\hline
       191 & 24556463604287900 &      14,30 &      11,66 &      12,94 &      12,43 &      13,14 &      11,92 &        253 & 24556563515848900 &       14,9 &       11,6 &       12,9 &       12,4 &       13,2 &       11,9 \\
\hline
       192 & 24556463633050400 &      14,29 &      11,64 &      12,94 &      12,45 &      13,17 &      11,92 &        254 & 24556563544656500 &       14,9 &       11,6 &         13 &       12,5 &       13,2 &       11,9 \\
\hline
       193 & 24556463661844000 &      14,25 &      11,65 &      12,94 &      12,44 &      13,17 &      11,90 &        255 & 24556563573489600 &       14,9 &       11,6 &       12,9 &       12,4 &       13,2 &       11,9 \\
\hline
       194 & 24556463690645800 &      14,26 &      11,66 &      12,95 &      12,45 &      13,17 &      11,90 &        256 & 24556563602315200 &       14,8 &       11,6 &       12,9 &       12,5 &       13,2 &       11,9 \\
\hline
       195 & 24556463719417500 &      14,29 &      11,66 &      12,95 &      12,44 &      13,16 &      11,91 &        257 & 24556563631170300 &       14,7 &       11,6 &         13 &       12,5 &       13,2 &       11,9 \\
\hline
       196 & 24556463748198400 &      14,26 &      11,65 &      12,94 &      12,44 &      13,17 &      11,90 &        258 & 24556563660010500 &       14,8 &       11,6 &       12,9 &       12,4 &       13,2 &       11,9 \\
\hline
       197 & 24556463776962100 &      14,27 &      11,65 &      12,94 &      12,45 &      13,16 &      11,90 &        259 & 24556563689034300 &       14,8 &       11,6 &         13 &       12,5 &       13,2 &       11,9 \\
\hline
       198 & 24556463805736300 &      14,28 &      11,65 &      12,94 &      12,44 &      13,14 &      11,89 &        260 & 24556563717838500 &       14,7 &       11,6 &         13 &       12,5 &       13,2 &       11,9 \\
\hline
       199 & 24556463834475500 &      14,25 &      11,66 &      12,91 &      12,45 &      13,15 &      11,90 &        261 & 24556563746656500 &       14,7 &       11,6 &         13 &       12,5 &       13,2 &       11,9 \\
\hline
       200 & 24556463863246000 &      14,29 &      11,65 &      12,95 &      12,45 &      13,15 &      11,90 &        262 & 24556563775566200 &       14,6 &       11,6 &         13 &       12,4 &       13,2 &       11,9 \\
\hline
       201 & 24556463892091600 &      14,27 &      11,66 &      12,95 &      12,45 &      13,16 &      11,90 &        263 & 24556563828103000 &       14,4 &       11,6 &       12,9 &       12,5 &       13,2 &       11,9 \\
\hline
       202 & 24556463920827400 &      14,27 &      11,64 &      12,93 &      12,46 &      13,17 &      11,89 &        264 & 24556563856946600 &       14,4 &       11,6 &       12,9 &       12,5 &       13,2 &       11,9 \\
\hline
       203 & 24556463949646600 &      14,31 &      11,66 &      12,94 &      12,44 &      13,14 &      11,90 &        265 & 24556563885721700 &       14,4 &       11,6 &         13 &       12,5 &       13,2 &       11,9 \\
\hline
       204 & 24556463978431000 &      14,33 &      11,66 &      12,95 &      12,46 &      13,16 &      11,89 &        266 & 24556563914558400 &       14,4 &       11,6 &         13 &       12,5 &       13,2 &       11,9 \\
\hline
       205 & 24556464007197800 &      14,28 &      11,66 &      12,93 &      12,44 &      13,17 &      11,91 &        267 & 24556563943384600 &       14,3 &       11,6 &         13 &       12,5 &       13,2 &       11,9 \\
\hline
       206 & 24556464036027400 &      14,28 &      11,65 &      12,93 &      12,47 &      13,13 &      11,90 &        268 & 24556563972191000 &       14,2 &       11,6 &         13 &       12,5 &       13,2 &       11,9 \\
\hline
       207 & 24556464064780500 &      14,32 &      11,65 &      12,95 &      12,47 &      13,13 &      11,91 &        269 & 24556564001077500 &       14,2 &       11,6 &       12,9 &       12,5 &       13,2 &       11,9 \\
\hline
       208 & 24556464093585800 &      14,34 &      11,67 &      12,94 &      12,44 &      13,13 &      11,89 &        270 & 24556564029828300 &       14,2 &       11,6 &       12,9 &       12,5 &       13,2 &       11,9 \\
\hline
       209 & 24556464122369000 &      14,29 &      11,66 &      12,96 &      12,45 &      13,14 &      11,90 &        271 & 24556564058639400 &       14,2 &       11,6 &       12,9 &       12,5 &       13,2 &       11,9 \\
\hline
       210 & 24556464151158100 &      14,34 &      11,64 &      12,96 &      12,43 &      13,13 &      11,90 &        272 & 24556564087514200 &       14,1 &       11,6 &       12,9 &       12,5 &       13,2 &       11,9 \\
\hline
       211 & 24556464179974900 &      14,36 &      11,65 &      12,95 &      12,45 &      13,14 &      11,90 &        273 & 24556564116356700 &       14,2 &       11,6 &       12,9 &       12,5 &       13,2 &       11,9 \\
\hline
       212 & 24556464255725400 &      14,46 &      11,64 &      12,97 &      12,43 &      13,18 &      11,91 &        274 & 24556564145191000 &       14,1 &       11,6 &       12,9 &       12,5 &       13,2 &       11,9 \\
\hline
       213 & 24556464284522600 &      14,45 &      11,65 &      12,94 &      12,44 &      13,19 &      11,91 &        275 & 24556564174032200 &       14,2 &       11,6 &       12,9 &       12,5 &       13,2 &       11,9 \\
\hline
       214 & 24556464313343400 &      14,41 &      11,66 &      12,94 &      12,44 &      13,11 &      11,91 &        276 & 24556564202928000 &       14,2 &       11,6 &       12,9 &       12,5 &       13,2 &       11,9 \\
\hline
       215 & 24556464342094200 &      14,34 &      11,65 &      12,94 &      12,44 &      13,16 &      11,91 &        277 & 24556564231788100 &       14,2 &       11,7 &       12,9 &       12,5 &       13,2 &       11,9 \\
\hline
       216 & 24556464370820700 &      14,44 &      11,67 &      12,94 &      12,42 &      13,17 &      11,90 &        278 & 24556564260599300 &       14,2 &       11,6 &         13 &       12,5 &       13,2 &       11,9 \\
\hline
       217 & 24556464399573800 &      14,41 &      11,65 &      12,94 &      12,44 &      13,15 &      11,90 &        279 & 24556564289420800 &       14,2 &       11,7 &       12,9 &       12,5 &       13,2 &       11,9 \\
\hline
       218 & 24556464428343500 &      14,46 &      11,66 &      12,94 &      12,43 &      13,15 &      11,90 &        280 & 24556564318225100 &       14,2 &       11,6 &         13 &       12,5 &       13,2 &       11,9 \\
\hline
       219 & 24556464457294900 &      14,42 &      11,65 &      12,94 &      12,43 &      13,13 &      11,90 &        281 & 24556564347026900 &       14,2 &       11,7 &       12,9 &       12,5 &       13,2 &       11,9 \\
\hline
       220 & 24556464486107100 &      14,47 &      11,65 &      12,94 &      12,46 &      13,14 &      11,90 &        282 & 24556564375926100 &       14,2 &       11,6 &         13 &       12,5 &       13,2 &       11,9 \\
\hline
       221 & 24556464514906700 &      14,46 &      11,66 &      12,95 &      12,45 &      13,13 &      11,89 &        283 & 24556564404780000 &       14,2 &       11,6 &       12,9 &       12,5 &       13,2 &       11,9 \\
\hline
       222 & 24556464572396800 &      14,45 &      11,64 &      12,93 &      12,43 &      13,16 &      11,91 &        284 & 24556564433609700 &       14,3 &       11,6 &         13 &       12,5 &       13,2 &       11,9 \\
\hline
       223 & 24556562624475600 &      14,94 &      11,64 &      12,93 &      12,46 &      13,17 &       11,9 &        285 & 24556564462423100 &       14,3 &       11,7 &       12,9 &       12,5 &       13,2 &       11,9 \\
\hline
       224 & 24556562653296700 &      14,88 &      11,63 &      12,94 &      12,45 &      3,161 &      11,91 &        286 & 24556564491219200 &       14,2 &       11,6 &       12,9 &       12,5 &       13,2 &       11,9 \\
\hline
       225 & 24556562682046400 &      14,93 &      11,64 &      12,93 &      12,46 &      13,15 &      11,91 &        287 & 24556564520023300 &       14,3 &       11,6 &         13 &       12,5 &       13,2 &       11,9 \\
\hline
       226 & 24556562710837800 &      14,85 &      11,64 &      12,94 &      12,46 &      13,17 &      11,91 &        288 & 24556564548884200 &       14,3 &       11,7 &         13 &       12,5 &       13,2 &       11,9 \\
\hline
       227 & 24556562739659100 &      14,84 &      11,64 &      12,95 &      12,46 &      13,16 &       11,9 &        289 & 24556564577710500 &       14,3 &       11,6 &       12,9 &       12,5 &       13,2 &       11,9 \\
\hline
       228 & 24556562768391400 &      14,85 &      11,64 &      12,94 &      12,46 &      13,16 &       11,9 &        290 & 24556564606509900 &       14,3 &       11,6 &         13 &       12,5 &       13,2 &       11,9 \\
\hline
       229 & 24556562797152600 &      14,85 &      11,64 &      12,95 &      12,46 &      13,17 &       11,9 &        291 & 24556564635374700 &       14,3 &       11,6 &       12,9 &       12,5 &       13,2 &       11,9 \\
\hline
       230 & 24556562825894200 &      14,82 &      11,64 &      12,94 &      12,45 &      13,16 &      11,91 &        292 & 24556564664202000 &       14,3 &       11,7 &       12,9 &       12,5 &       13,2 &       11,9 \\
\hline
       231 & 24556562854765000 &      14,84 &      11,64 &      12,95 &      12,46 &      13,16 &       11,9 &        293 & 24556564693538100 &       14,3 &       11,6 &       12,9 &       12,5 &       13,2 &       11,9 \\
\hline
       232 & 24556562883505500 &      14,81 &      11,64 &      12,94 &      12,45 &      13,14 &      11,91 &            &            &            &            &            &            &            &            \\
\hline
\end{tabular}  
\end{table}
\begin{table}
\caption{Comparison of AM Cnc parameters with published data} 
\begin{tabular}{|cccccr|}
\hline
{\bf Star} & {\bf Type} &             \multicolumn{ 4}{c|}{{\bf Period (d)}} \\
\hline
           &            &            &            &            &            \\
\hline
\multicolumn{ 1}{|c}{{\bf AM Cnc}} & \multicolumn{ 1}{c}{{\bf RRAB}} & {\bf This Work} & {\bf GCVS} & {\bf Konkoly Observatory\cite{Konkoly:ref1}} & {\bf AAVSSO} \\
\multicolumn{ 1}{|c}{{\bf }} & \multicolumn{ 1}{c}{{\bf }} &   0.559233 &   0.557615 &    0.55803 &  0.5580015 \\
\hline
\end{tabular}  
\end{table}
\clearpage
\newpage
\includegraphics[width=0.5\textwidth{}]{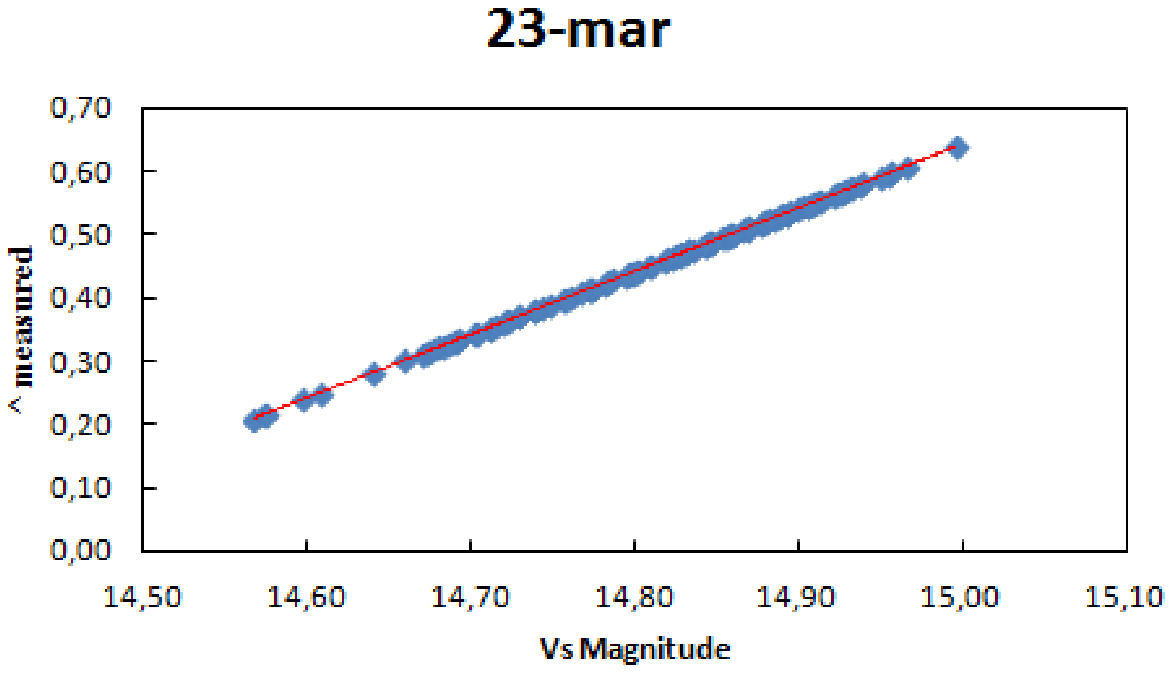} 
\includegraphics[width=0.5\textwidth{}]{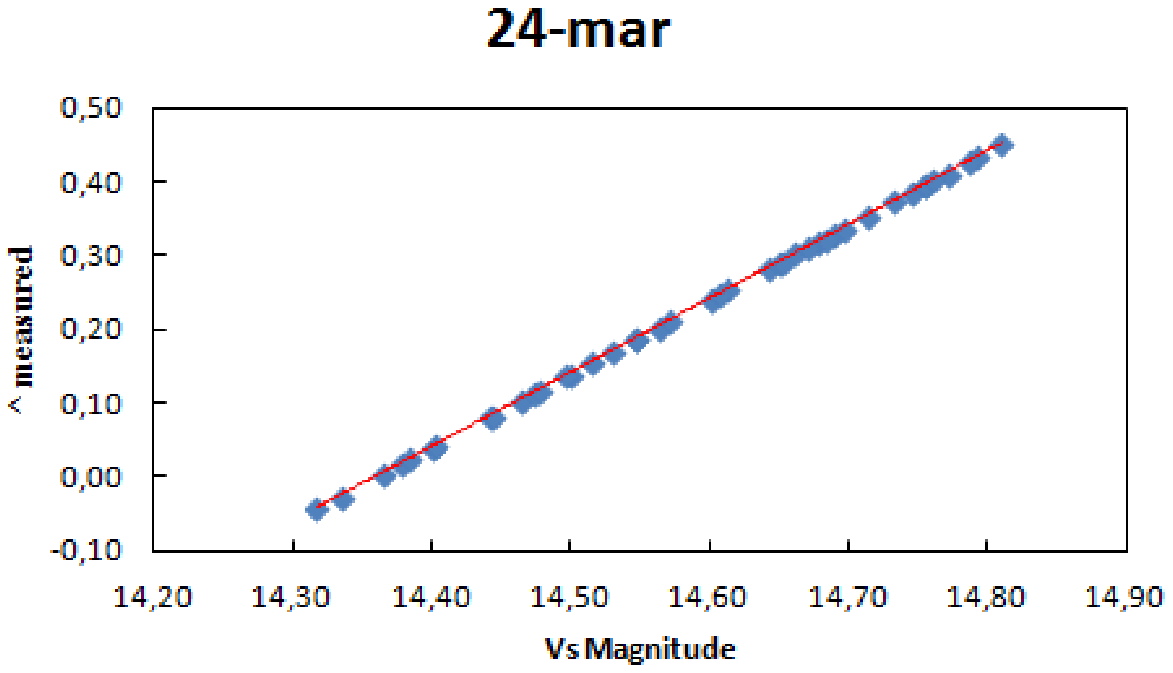}
\includegraphics[width=0.5\textwidth{}]{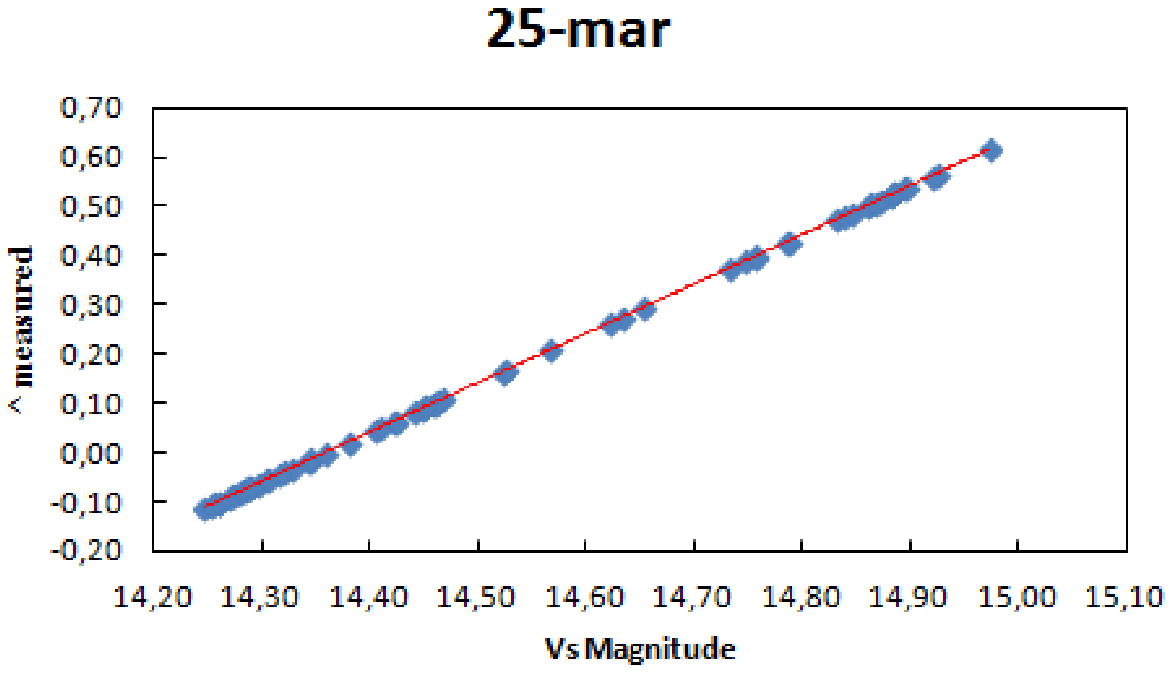} 
\includegraphics[width=0.5\textwidth{}]{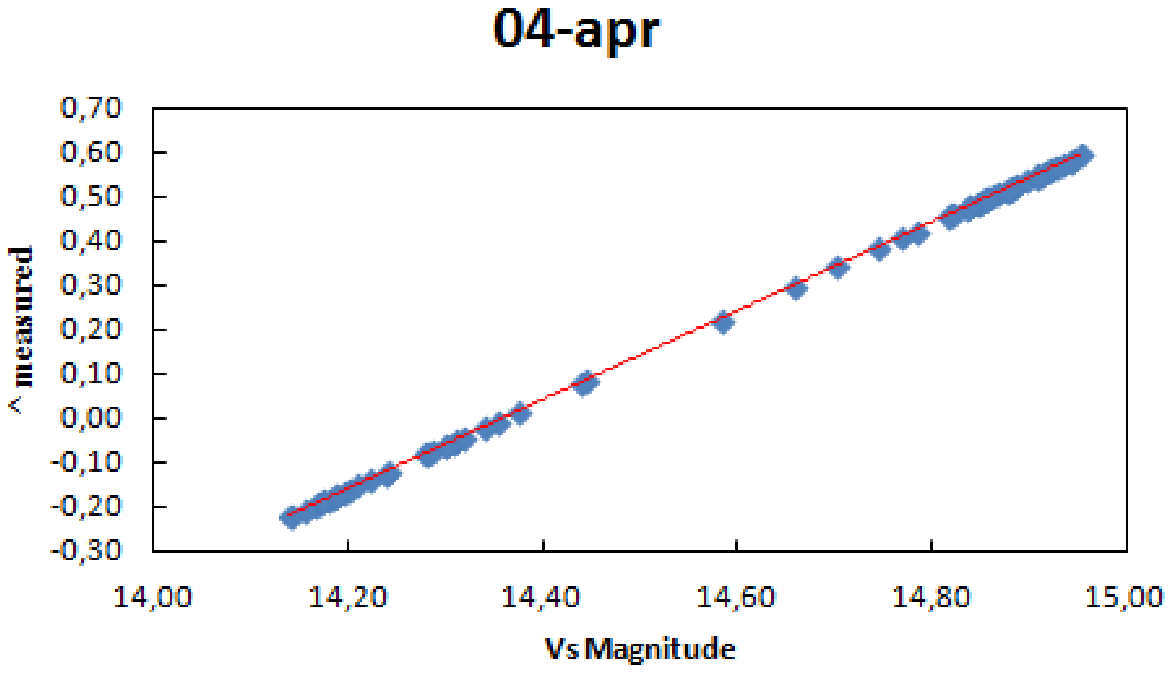} \\
\textbf{{\scriptsize Figure 3-4-5-6: Variation between instrumental measurement and value stellar catalog for each observing session}} \\

\end{document}